\documentclass[aps,preprint,nofootinbib]{revtex4-1}
\usepackage[utf8]{inputenc}
\usepackage{amsfonts}
\usepackage{color}

\usepackage{amssymb}
\usepackage{amsmath}
\usepackage{stmaryrd}
\usepackage{latexsym}

\usepackage{pbox}
\usepackage{xcolor}
\definecolor{myurlcolor}{HTML}{08457E}
\definecolor{mylinkcolor}{HTML}{2A52BE}
\definecolor{mycitecolor}{HTML}{E30022}

\usepackage{float}
\usepackage[colorlinks, linkcolor=mylinkcolor, citecolor=mycitecolor, urlcolor=myurlcolor, linktocpage=true]{hyperref}

\def\equationautorefname~#1\null{(#1)\null}
\def\tableautorefname~#1\null{(#1)\null}
\def\figureautorefname~#1\null{(#1)\null}
\def\sectionautorefname~#1\null{(#1)\null}

\let\origref\autoref
\def\autoref#1{\textbf{\origref{#1}}}

\let\origcite\cite
\def\cite#1{\textbf{\origcite{#1}}}

\usepackage{multirow}

\usepackage{tikz}

\usepackage{titlesec}
\titleformat*{\section}{\centering\small\bfseries\scshape}
\titleformat*{\subsection}{\small\bfseries\scshape}
\titleformat*{\subsubsection}{\small\bfseries\scshape}

\newcommand{\be}{\begin{equation}}
\newcommand{\ee}{\end{equation}}
\newcommand{\bea}{\begin{eqnarray}}
\newcommand{\eea}{\end{eqnarray}}
\newcommand{\benn}{\begin{eqnarray*}}
\newcommand{\eenn}{\end{eqnarray*}}

\newcommand{\Ct}{@{\hspace{1em}} c @{\hspace{1em}}}
\newcommand{\Lt}{@{\hspace{1em}} l @{\hspace{1em}}}

\def\bse{\begin{subequations}}%
\def\ese{\end{subequations}}%

\newcommand{\sfrac}[2]{\dfrac{\,#1\,}{\,#2\,}}

\let\oldsqrt\sqrt
\def\sqrt{\mathpalette\DHLhksqrt}
\def\DHLhksqrt#1#2{%
	\setbox0=\hbox{$#1\oldsqrt{#2\,}$}\dimen0=\ht0
	\advance\dimen0-0.4\ht0
	\setbox2=\hbox{\vrule height\ht0 depth -\dimen0}%
	{\box0\lower0.4pt\box2}}

\def\jcap{Journal of Cosmology and Astroparticle Physics}
\def\prd{Physical Review D}

\begin{document}

\title{Dynamical System Analysis \\ of a \\ Five-Dimensional Cosmological Model}

\author{A. Sava{\c s} Arapo{\u g}lu}
\email{arapoglu@itu.edu.tr}

\author{Ezgi Canay}
\email{ezgicanay@itu.edu.tr}

\author{A. Emrah Y{\"u}kselci}
\email{yukselcia@itu.edu.tr}
\affiliation{Istanbul Technical University, Department of Physics, 34469 Maslak, Istanbul, Turkey \vspace{2cm}}

\begin{abstract}
	A five-dimensional cosmological model including a single perfect fluid is studied in the framework of dynamical system analysis. All the critical points of the system are listed with their stability properties and some representative phase diagrams are explicitly shown. It is found that the stabilization of extra dimension is possible and the observed flatness of the three-dimensional space is provided for certain ranges of the equation of state parameter of the fluid. The model suggested here can be considered as a simplified model for examining the possible effects of the extra dimensions in the early universe.	
\end{abstract}

\maketitle

\raggedbottom

\section{INTRODUCTION}
The idea of extra dimensions dates back to the original proposals of Kaluza and Klein (\cite{Kaluza,Klein}, and see \cite{goenner04} for a review) to unify the two known successful classical theories of their time, electromagnetic theory with Einstein's general relativity by introducing a compact fifth dimension. After a long interlude the revival of interest in extra dimensions started at seventies with the search of fundamental theories of matter and interactions like string theory formulated in more than three space dimensions. The presence of the extra dimensions in these theories, however, are forced by mathematical consistency unlike the original proposal of Kaluza-Klein. Another attempt with the use of extra dimensions is the so called brane-world models (\cite{rs1,rs2,add,dgp,maartens}) motivated by the string/M-theory to solve particularly the hierarchy problem based on the possibility of the existence of non-compact extra dimensions. The observation of the accelerated expansion of the universe (\cite{perlmutter,riess}) is another subject attracting much attention, and the machinery of the extra dimensions are used in this context also (\cite{town03,moham,pahwa11,appel01,bringmann1}).

Dynamical system approach is a powerful mathematical method to extract information about the global dynamics of a system whose evolution can be described by an autonomous system of differential equations. Equations describing the dynamics of the universe can also be cast into the form of a dynamical system by suitable choice of parameters (See \cite{dynreview} and references therein). This method thus provides a tool for investigating cosmological models through the use of their phase space by considering critical points, their stability, attractor behaviour, etc. This kind of analysis can help one see whether the model under consideration is capable of describing the observed properties of the universe and may lead to restrictions for the free parameters of the model and even eliminate some models.

In this paper we will apply the methods of the dynamical system analysis to a five-dimensional cosmological model that contains a single perfect fluid. It can be considered as a toy model to be applied to the early universe because it is more reasonable to consider the possible effects of the extra dimensions in this epoch rather than in late-time dynamics. We particularly examine whether the stabilization of the extra dimension (\cite{wald-carroll02,bringmann2}) is possible and the observed flatness of the three-dimensional space can be provided regardless of its initial value in this setting.

The paper is organized as follows: in Section 2, we present the set-up of the model, field equations, and the autonomous system. In Section 3, the critical points and the conditions for their stability are presented together with their physical meanings. In Section 4, some cosmologically interesting and relevant cases are considered, and the Section 5 is for the concluding remarks.

\section{AN AUTONOMOUS SYSTEM FOR A HOMOGENEOUS UNIVERSE \\ IN $(1+3+1)$-DIMENSIONS}
We consider a $(1+3+1)$-dimensional anisotropic universe described by the FRW-type metric
\begin{equation}
	ds^2 = -dt^2 + a^2(t) \bigg[ \sfrac{dr^2}{1-kr^2} + r^2 (d\theta^2 + sin^2\theta \, d\phi^2) \bigg] + b^2(t) \, dy^2
	\label{eq:metric}
\end{equation}
in which $a(t)$ is the scale factor of the 3-dimensional space, $y$ and $b(t)$ are the coordinate and the scale factor of the extra spatial dimension, respectively. The topology of the extra dimension is taken as $S^1$. In this setting we keep the curvature of space to see its role on the higher dimensional model (if any).

Assuming also that the dynamics of such a universe is governed by the generalized 4-dimensional Einstein gravity, the field equations are of the from 
\begin{equation}
	G_{AB} = \kappa_5 \, T_{AB} \: ,
	\label{eq:fieldeqs}
\end{equation}
where $A,B=1,2,3,4,5$ and $\kappa_5$ is the five dimensional gravitational constant which is related to the four dimensional one in the case of a compactified extra dimension with radius $r_{5}$ through 
\begin{equation}
	\kappa_5 =\frac{\kappa}{2\pi r_{5}}.
\end{equation}

We consider a single perfect fluid which is described by an energy-momentum tensor of the form
\begin{equation}
	T^{A}_{\phantom A B} = diag[-\rho, p, p, p, p_5]
	\label{eq:enmom}
\end{equation}
where the pressure along the three conventional directions is $p=\omega \rho$ and along the extra dimension $p_5=\omega_5 \rho$, with $\omega$ and $\omega_5$ as the corresponding equation of state (EoS) parameters.	

Using Eq.\autoref{eq:metric} together with Eq.\autoref{eq:enmom} in Eq.\autoref{eq:fieldeqs}, we derive the Friedmann equations as follows
\begin{equation}
	\begin{aligned}
		3\sfrac{\dot{a}^2}{a^2}+3\sfrac{\dot{a}}{a}\sfrac{\dot{b}}{b}+3\sfrac{k}{a^2}&=\kappa_5 \rho \\[1mm]
		-2\sfrac{\ddot{a}}{a}-\sfrac{\dot{a}^2}{a^2}-2\sfrac{\dot{a}}{a}\sfrac{\dot{b}}{b}-\sfrac{\ddot{b}}{b}-\sfrac{k}{a^2}&=\kappa_5 (w\rho) \\[1mm]
		-3\sfrac{\ddot{a}}{a}-3\sfrac{\dot{a}^2}{a^2}-3\sfrac{k}{a^2}&=\kappa_5 (w_5\rho) \: .
	\end{aligned}
	\label{eq:friedmanneqs}
\end{equation}

Defining $\dot{a}/a \equiv H$ and $\dot{b}/b \equiv h$ as the Hubble parameters of the three dimensional observed space and that of the extra dimension, the field equations can be cast into the form	
\begin{equation}
	\begin{aligned}
		3H^2+3Hh+3\sfrac{k}{a^2}&=\kappa_5\rho \\[1mm]
		-2\dot{H}-3H^2-2Hh-\dot{h}-h^2-\frac{k}{a^2}&=\kappa_5 (w\rho) \\[1mm]
		-3\dot{H}-6H^2-3\sfrac{k}{a^2}&=\kappa_5 (w_5\rho) \: .\\[1mm]
	\end{aligned}
	\label{eq:Hfred}
\end{equation}
Dividing both sides of the first equation in Eq.\autoref{eq:Hfred} by $3H^2$ and switching to dimensionless parameters $\Omega_\rho$, $\Omega_c$ and $\Omega_h$, we can write
\begin{equation}
	1+ \Omega_h + \Omega_c = \Omega_\rho
	\label{eq:dimpar}
\end{equation}
in which
\begin{equation}
	\Omega_{\rho}=\sfrac{\kappa_5\rho}{3H^2} \: ,
	\qquad
	\Omega_h=\sfrac{h}{H} \: ,
	\qquad
	\Omega_c=\sfrac{k}{a^2H^2} \: .
	\label{eq:dimpardef}
\end{equation}
The corresponding set of autonomous equations encoding the time evolution of the system is obtained from Eq.\autoref{eq:Hfred} and  Eq.\autoref{eq:dimpardef} as
\begin{equation}
	\begin{aligned}
		\Omega'_\rho &=\Omega_\rho \Big[ (1-3w) + 2w_5 \, \Omega_\rho-(1+w_5)\Omega_h \Big] \\[1mm]
		\Omega'_h &= 1 - \Omega_h^2 + \Omega_c + (2w_5-3w)\Omega_\rho + (w_5 \, \Omega_\rho + \Omega_c) \Omega_h \\[1mm]
		\Omega'_c &= 2\Omega_c \Big[ \Omega_c + w_5 \, \Omega_\rho + 1 \Big]
	\end{aligned}
	\label{eq:dyn_sys}
\end{equation}
where the prime denotes derivative with respect to $N\equiv \ln a$. Making use of Eq.\autoref{eq:dimpar} we reduce the number of independent variables to two. In order to focus on the behavior of $\Omega_h$ and $\Omega_c$, we eliminate the dimensionless energy density parameter $\Omega_\rho$ of the five-dimensional fluid and obtain the autonomous system as
\begin{equation}
	\begin{aligned}
		\Omega'_c &= 2 \Omega_c \Big[ \Omega_c + w_5 (1 + \Omega_c + \Omega_h) + 1 \Big] \\[1mm]
		\Omega'_h &= 1 - \Omega_h^2 + \Omega_c(1 + \Omega_h) + (1 + \Omega_c + \Omega_h) \Big[ (2 + \Omega_h)w_5 - 3 w \Big] \: .
	\end{aligned}
	\label{eq:dynsys}
\end{equation}

\section{CRITICAL POINTS AND STABILITY ANALYSIS}
The autonomous system in Eq.\autoref{eq:dynsys} has four critical points listed in Table \autoref{tab:1}.

Before exploring the physical interpretations of these critical points, we introduce an extended version of bifurcation diagrams that are commonly used in dynamical system analysis.

\def\arraystretch{1.5}
\begin{table*}[t]
	\scriptsize
	\centering
	%\captionsetup{margin=1cm}
	\caption{Critical points of autonomous system given in Eq.\eqref{eq:dynsys}.}
	\label{tab:1}
	\hspace*{-0.3cm}
	\begin{tabular}{\Ct|\Ct|\Ct|\Lt|\Ct|\Lt} \hline
		\# & \boldmath $\Omega_c$ & \boldmath $\Omega_h$ & \textbf{Eigenvalues} & \textbf{Condition} & \textbf{Character} \\[.3mm] \hline \hline
		
		\multirow{2}{*}{A} & \multirow{2}{*}{$0$} & \multirow{2}{*}{$-1$} & $\lambda_1 = 2$ & $w_5>3w-2$ & Unstable node \\ 
		& & & $\lambda_2 = w_5-3w+2$ & $w_5<3w-2$ & Saddle node \\ \hline
		
		\multirow{2}{*}{B} & \multirow{2}{*}{$-1$} & \multirow{2}{*}{$0$} & $\lambda_1 = -2$ & $w>-1/3$ & Stable node \\ 
		& & & $\lambda_2 = -3w-1$ & $w<-1/3$ & Saddle node \\ \hline
		
		\multirow{3}{*}{C} & \multirow{3}{*}{$0$} & \multirow{3}{*}{$\sfrac{2w_5-3w+1}{1-w_5}$} & \multirow{2}{*}{$\lambda_1 =  \sfrac{2w_5(w_5-3w+1)+2}{1-w_5}$} & $\lambda_1 > 0 \wedge \lambda_2 > 0 $ & Unstable node \\ 
		& & & & $\lambda_1 \cdot \lambda_2 < 0 $ & Saddle node \\
		& & & $\lambda_2 = -w_5+3w-2$ & $\lambda_1 < 0 \wedge \lambda_2 < 0 $ & Stable node \\ \hline
		
		\multirow{4}{*}{D} & \multirow{4}{*}{$-\sfrac{w_5(w_5-3w+1)+1}{(w_5+1)^2}$} & \multirow{4}{*}{$-\sfrac{3w+1}{w_5+1}$} & \multirow{2}{*}{$\lambda_1 = g(w,w_5) - \sqrt{f(w,w_5)}$} & $g < 0 \wedge f < 0 $ & Stable spiral \\ 
		& & & & $\lambda_1 < 0 \wedge \lambda_2 < 0 \wedge f \geq 0 $ & Stable node \\
		& & & \multirow{2}{*}{$\lambda_2 = g(w,w_5) + \sqrt{f(w,w_5)}$} & $\lambda_1 \cdot \lambda_2 < 0 \wedge f \geq 0  $ & Saddle node \\
		& & & & $g > 0 \wedge f < 0 $ & Unstable spiral \\ \hline\hline
		
		\multicolumn{6}{c}{\multirow{3}{*}{$f(w,w_5) = \sfrac{3\big[4(w_5)^2(2w+1)-4w_5(6w^2+w-1)+3(w+1)^2\big]}{4(w_5+1)^2}$ \hspace{1.5cm} $g(w,w_5) = \sfrac{3w-2w_5-1}{2(w_5+1)}$}} \\
		\multicolumn{6}{c}{} \\
		\multicolumn{6}{c}{} \\
		
		\hline\hline
	\end{tabular}
	\vspace{2mm}
\end{table*}

Bifurcation diagrams appear as crucial tools that demonstrate how control parameters (if, available) in a system affect the stability of fixed points and consequent solution curves. Such diagrams become relevant to our work when we look at Table \autoref{tab:1} for the conditions that determine the characteristics of points A, B, C and D. Conventionally, they are given in the form of critical point versus control parameter plots in order to ensure clear representation of the relation pattern in between. Since there are two independent control parameters in our model that are $w$ and $w_5$, we obviously need three dimensional versions of these bifurcation diagrams. However, $w-w_5$ plane with color-coded representation in Fig.\autoref{fig:stabli} suffices to manifest the dynamics of our system and for the purpose of simplicity, we may then avoid the usage of rather complicated three dimensional diagrams in this context. Accordingly, the generic notion of bifurcation point is replaced by two dimensional objects which we call \textit{bifurcation curves}. Inspired from the two dimensional bifurcation diagrams, they correspond to curves that form the border lines of different stability characters of the fixed points.

In the light of above considerations, we investigate the four critical points in the model as listed in Table \autoref{tab:1}. The main interest here is in the physical interpretation of the fixed points; thus, each will be studied in terms of its cosmological aspects in detail. In order to avoid loss of generality, we impose no constraints on the values of $\Omega_h$ and $\Omega_c$ initially but instead we study the evolution of the system under two main cases, namely $\Omega_c < 0$ and $\Omega_c > 0$, the reason of which will be clear in the next section with the discussion of the flat universe case, i.e. $\Omega_c = 0$. While investigating the appropriate EoS parameters, we mainly restrict ourselves to the range $-1\leq w,w_5 \leq 1$ for compatibility purposes.

\begin{itemize}
	\setlength\itemsep{.5mm}
	\item \textbf{\small Point A} corresponds to a flat universe and it is either an unstable node or a saddle node depending on the values of EoS parameters given under stability conditions in Table \autoref{tab:1}. Its stability is interchangeable with Point C via transcritical bifurcation. Bifurcation curve here obeys the equation $w_5 = 3w - 2$ as seen in Fig.\autoref{fig:pointA}. 
	
	Physically, the state corresponds to a moment dominated by the contraction of the fifth dimension with $\Omega_h=-1$, $\Omega_c=0$, and $\Omega_\rho=0$. In a recent work (\cite{UED2}) it appears as one of the Kasner-type solutions to the set of differential equations governing the time evolution of both Hubble parameters in a flat and five dimensional universe. Authors obtain this result under the additional constraint $h(t)=cH(t)$ for some constant $c$ which corresponds to $\Omega_h = c$ in our case.
	
	\begin{figure*}[t]
		\centering
		
		\begin{tabular}{@{}c@{}}
			\includegraphics[width=.24\linewidth]{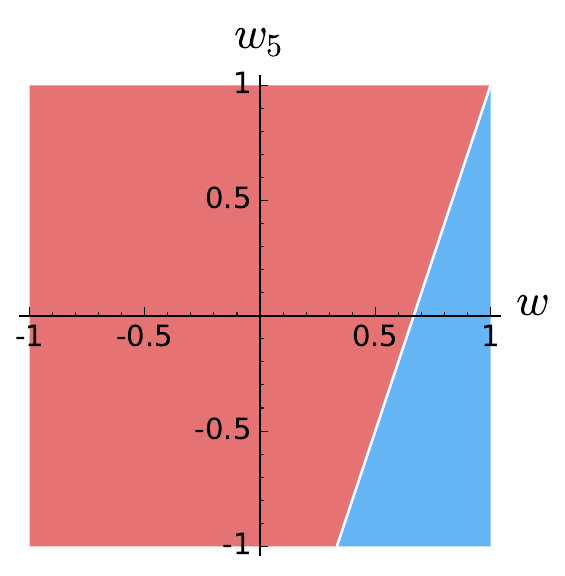}\\
			\small \textbf{(a)} Point A
			\label{fig:pointA}
		\end{tabular}
		\begin{tabular}{@{}c@{}}
			\includegraphics[width=.24\linewidth]{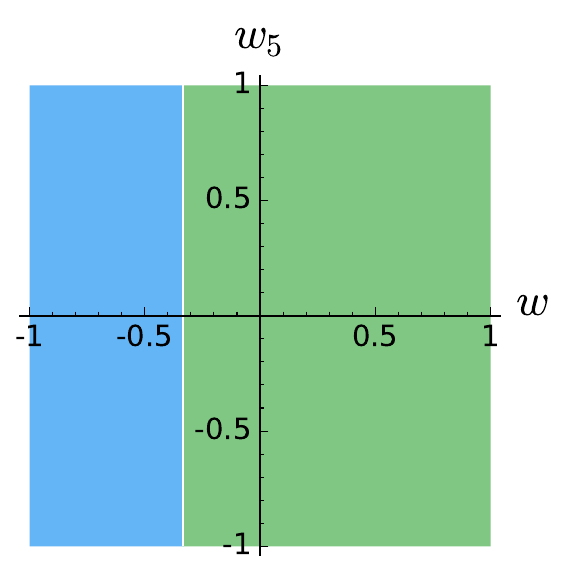}\\
			\small \textbf{(b)} Point B
			\label{fig:pointB}
		\end{tabular}
		\begin{tabular}{@{}c@{}}
			\includegraphics[width=.24\linewidth]{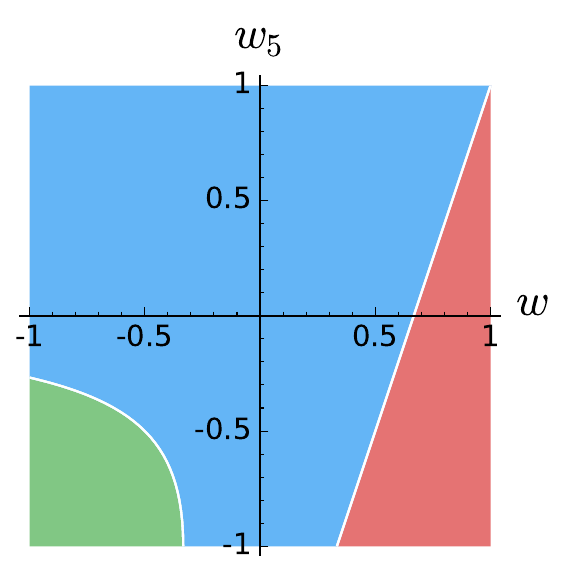}\\
			\small \textbf{(c)} Point C
			\label{fig:pointC}
		\end{tabular}
		\begin{tabular}{@{}c@{}}
			\includegraphics[width=.24\linewidth]{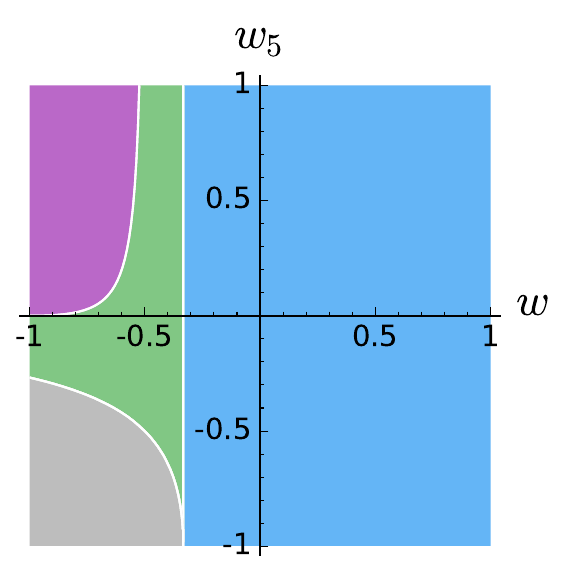}\\
			\small \textbf{(d)} Point D
			\label{fig:pointD}
		\end{tabular}
		
		\vspace{-0.5mm}
		
		\caption{Stability analysis of the critical points with respect to EoS parameters. White curves between regions represent bifurcation curves. Green : Stable node, Blue : Saddle node, Red : Unstable node, Purple : Stable spiral, Gray : Saddle node. There is no constraint on characters of points A, B and C. As for Point D, the gray region is valid for $\Omega_c > 0$, while the rest exists for $\Omega_c < 0\,$.}
		\label{fig:stabli}
	\end{figure*}
	
	\item \textbf{\small Point B} indicates a negatively curved universe over the entire plane in Fig.\autoref{fig:pointB} and its stability depends only on $w$ that is defined along the three spatial coordinates. For $w<-1/3$, it appears as a saddle node and consequently allows vanishing curvature in cases where Point C is the attractor. As also implied by the Friedmann equation, fluids with $w<-1/3$ can overcome curvature. The equation $w = -1/3$, therefore, represents again a transcritical bifurcation curve between points B and D.
	
	The patch with Point B as the attractor shown in Fig.\autoref{fig:pointB} well satisfies the stability condition with $\Omega_h=0$. However, one should be careful here because this also means our single component universe will evolve to a negatively curved space. There exists no analogous attractor to this point for the case of a closed universe.
	
	\item \textbf{\small Point C} is the only fixed point at which both the flatness condition and the stabilization of the extra dimension can be achieved simultaneously. The universe appears flat for all the EoS parameters due to the constant $\Omega_c$ given in Table \autoref{tab:1}, but exact stabilization occurs only on the line $2w_5-3w+1=0$ as also expressed in (\cite{bringmann1}) while solving field equations in the context of a static extra dimension with $\dot{b}=\ddot{b}=0$. Within the plane $-1\leq w,w_5 \leq 1$, Point C as an attractor can never satisfy $\Omega_h=0$. Therefore it is necessary to allow $w_5<-1$ and impose the condition $w<-1/3$ so that the system goes into a state with both stabilized extra dimension and vanishing curvature. Again, another Kasner-type solution\footnote{In (\cite{UED2}), $c_3=\frac{1-3w+2w_5}{1+(n-1)w-nw_5}$ where $c_3=\Omega_h$ in our case and $n=1$.} obtained in (\cite{UED2}) appears at this point as one of the states that may naturally be achieved along the time evolution of the autonomous system in Eq.\autoref{eq:dynsys}.
	
	This particular fixed point has two transcritical bifurcations one of which occurs on the curve $w_5(w_5-3w+1)+1=0$ with Point D and the other on the line $w_5 = 3w - 2$ with Point A.
	
	\item \textbf{\small Point D} is the most intriguing critical point in terms of its stability properties. As mentioned above, the first bifurcation associated with Point D occurs with Point C. The other appears in the form of Hopf bifurcation in which the eigenvalues of a critical point pass through the imaginary axis to produce periodic solutions. The equation $f(w_5,w) = 0$ specified in Table \autoref{tab:1} yields the bifurcation curve within Point D and the sign of $f(w_5,w)$ together with that of $g(w_5,w)$ serve to set the stability conditions for stable and unstable spirals. 
	
	Further analysis of the point shows that even though such behavior is allowed mathematically, it cannot be a proper physical attractor due to the fact that all $\Omega_h=0$ states lie on the bifurcation line $w=-1/3$ whereas the point may be the attractor of the system only for $w<-1/3$. For the special case of $w>1$ and $w_5>1$, that is outside the plane in Fig.\autoref{fig:pointD}, Point D becomes an unstable spiral.
\end{itemize}

\def\arraystretch{1.5}
\begin{table*}[t]
	\scriptsize
	\centering
	\caption{Cosmologically interesting cases listed together with the stability of points A, B, C and D at the corresponding locations on the $w,w_5$ plane. The stability properties written in bold are labeled forbidden for they allow $\Omega_h\geq0$.}
	\label{my-label}
	\begin{tabular}{c|c|c|c|c|c}
		& $\quad$ Quadrant $\quad$ & $\qquad$Point A$\qquad$  & $\qquad$Point B$\qquad$ & $\qquad$Point C$\qquad$ & $\qquad$Point D$\qquad$ \\ \hline\hline
		\multicolumn{1}{c|}{\multirow{5}{*}{\pbox{3cm}{Isotropic Fluid \\ $(w=w_5)$}}} & $First$ & \multirow{5}{*}{Unstable node} & \multirow{2}{*}{Stable node} & \multirow{3}{*}{\textbf{Saddle node}} & \multirow{2}{*}{Saddle node} \\ \cline{2-2}  
		\multicolumn{1}{c|}{}                  & \multirow{4}{*}{$Third$} & &  &  &  \\ \cline{4-4}  \cline{6-6}
		\multicolumn{1}{c|}{}                  &  & & \multirow{2}{*}{Saddle node} &  & \textbf{Stable node} \\ \cline{5-6}  
		\multicolumn{1}{c|}{}                  &  & &  & \multirow{2}{*}{\textbf{Stable node}} & None \\ \cline{4-4}\cline{6-6}  
		\multicolumn{1}{c|}{}                  &  & & None &  & \textbf{Saddle node} ($\Omega_c>0$) \\ \hline\hline
		\multicolumn{1}{c|}{\multirow{5}{*}{\pbox{3cm}{Stabilization of Extra Dimension \\ $(1+2w_5-3w=0)$}}} & $First$ & \multirow{5}{*}{Unstable node} & \multirow{3}{*}{Stable node} & \multirow{3}{*}{Saddle node} & \multirow{3}{*}{Saddle node} \\ \cline{2-2} 
		\multicolumn{1}{l|}{}                  & $Fourth$ &  &  &  &  \\ \cline{2-2} 
		\multicolumn{1}{l|}{}                  & $Third$ &  &  &  &  \\ \cline{2-2}\cline{4-6}
		\multicolumn{1}{l|}{}                  & \multirow{2}{*}{$Third$\footnote{Outside the zone $-1\leq w,w_5\leq 1$ for $w < -1/3$ and $w_5 < -1$}} &  & Saddle node & Stable node & None \\  \cline{4-6}
		\multicolumn{1}{l|}{}                  &  &  & None & Stable node & \textbf{Saddle node} ($\Omega_c>0$) \\ \hline\hline
		\multicolumn{1}{c|}{\multirow{3}{*}{\pbox{3cm}{Highly-Relativistic Fluid \\ $(w_5+3w=1)$}}} & $First$ &\multirow{2}{*}{ Unstable node} & \multirow{3}{*}{Stable node} & \multirow{2}{*}{Saddle node} & \multirow{3}{*}{Saddle node} \\ \cline{2-2} 
		\multicolumn{1}{l|}{}                  & \multirow{2}{*}{$Fourth$} &  &  &  &  \\ \cline{3-3}\cline{5-5}
		\multicolumn{1}{l|}{}                  &  & Saddle node &  & Unstable node &  \\ \hline\hline
	\end{tabular}
\end{table*}

\section{COSMOLOGICALLY INTERESTING SOLUTIONS}
In our five-dimensional model, we investigate the possibility of obtaining an epoch during which the extra-dimension reduces in size down to non-observable scales and gets stabilized which corresponds to $\Omega_h\!=\!0$. We try to find the correct combinations of $w$, $w_5$, i.e. the  appropriate five-dimensional fluid, that would allow the stabilization of the fifth dimension and prepare the background of standard cosmology.	

In order to point out the solutions of physical interest, we look at three main categories labeled as isotropic fluid, stabilization of extra dimension, and highly-relativistic fluid in Table \autoref{my-label} where each case is given together with the line it represents on the $w-w_5$ plane. Focusing on the attractors of these three categories, we describe the evolutionary patterns accessible through the dynamical analysis of higher dimensional field equations with the specific choice of variables in Eq.\autoref{eq:dimpardef}.

\subsection{Non-zero Pre-Inflationary Curvature}
\begin{figure*}[t]
	\centering
	
	\begin{tabular}{@{}c@{}}
		\includegraphics[width=.295\linewidth]{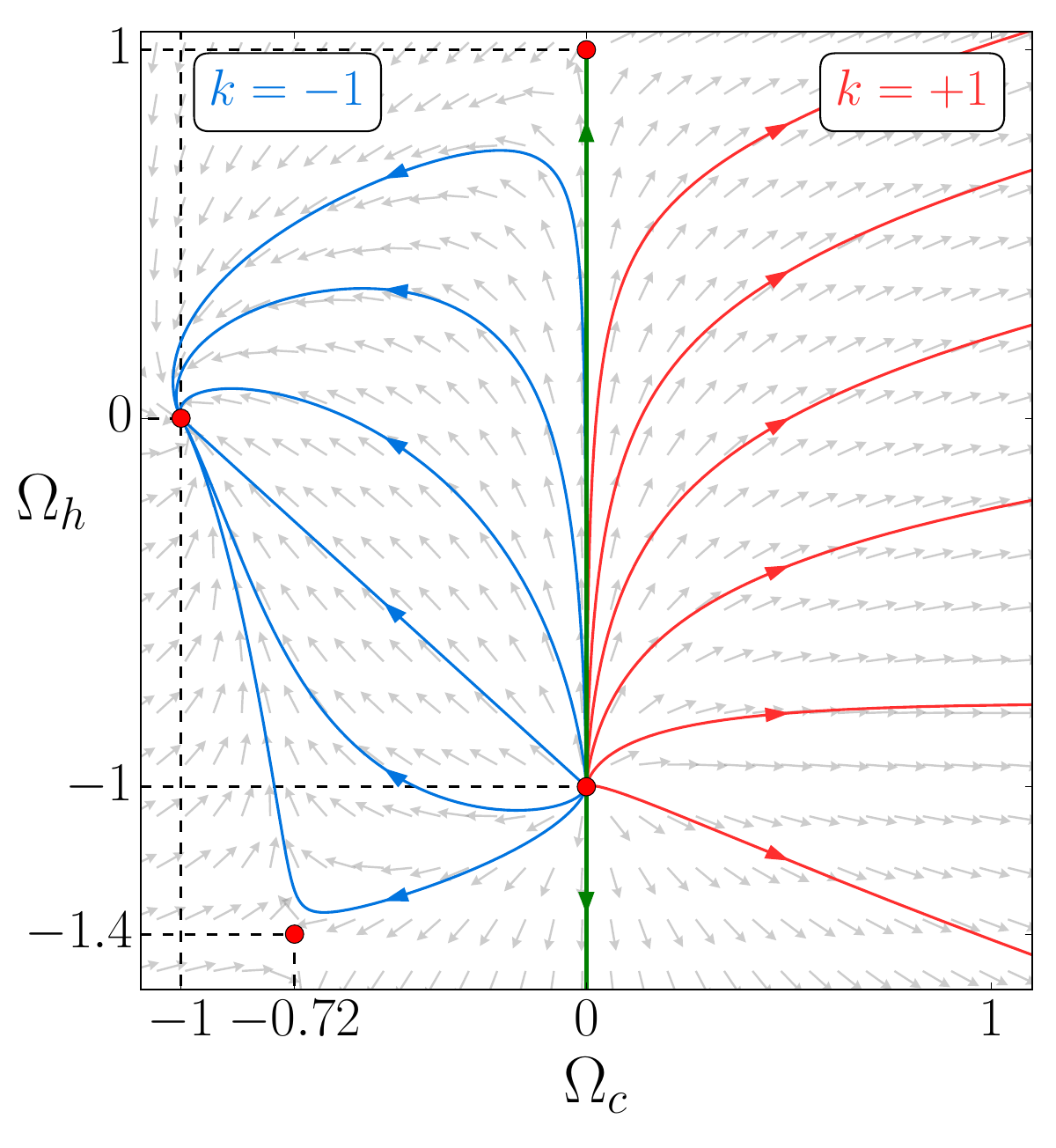} \\[-2mm]
		\small \textbf{(a)} $w=w_5=1/4$
		\label{fig:cur_ext_rad}
	\end{tabular}
	\begin{tabular}{@{}c@{}}
		\includegraphics[width=.32\linewidth]{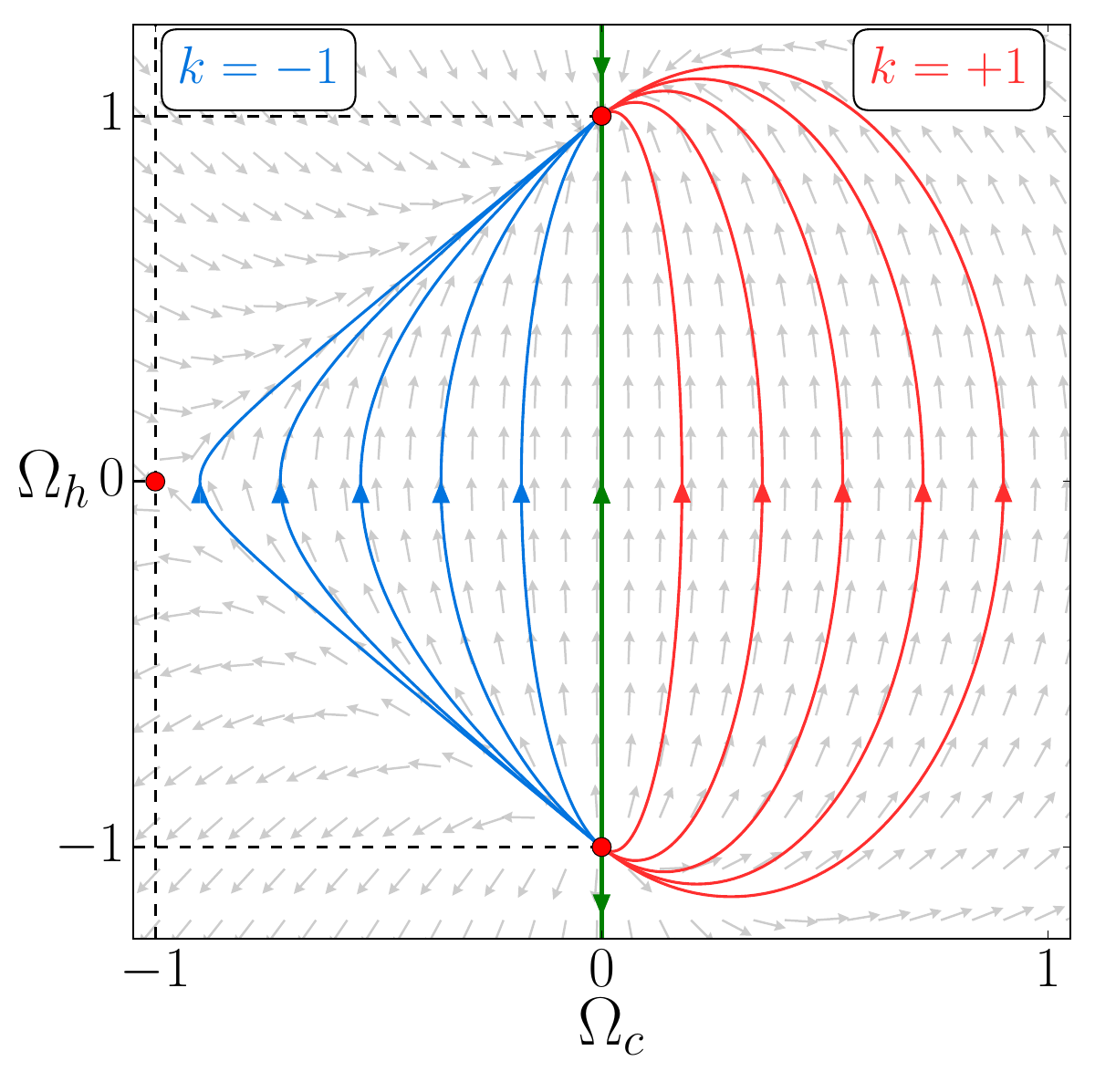} \\[-2mm]
		\small \textbf{(b)} $w=w_5=-1$
		\label{fig:cur_ext_lambda}
	\end{tabular}
	\begin{tabular}{@{}c@{}}
		\includegraphics[width=.36\linewidth]{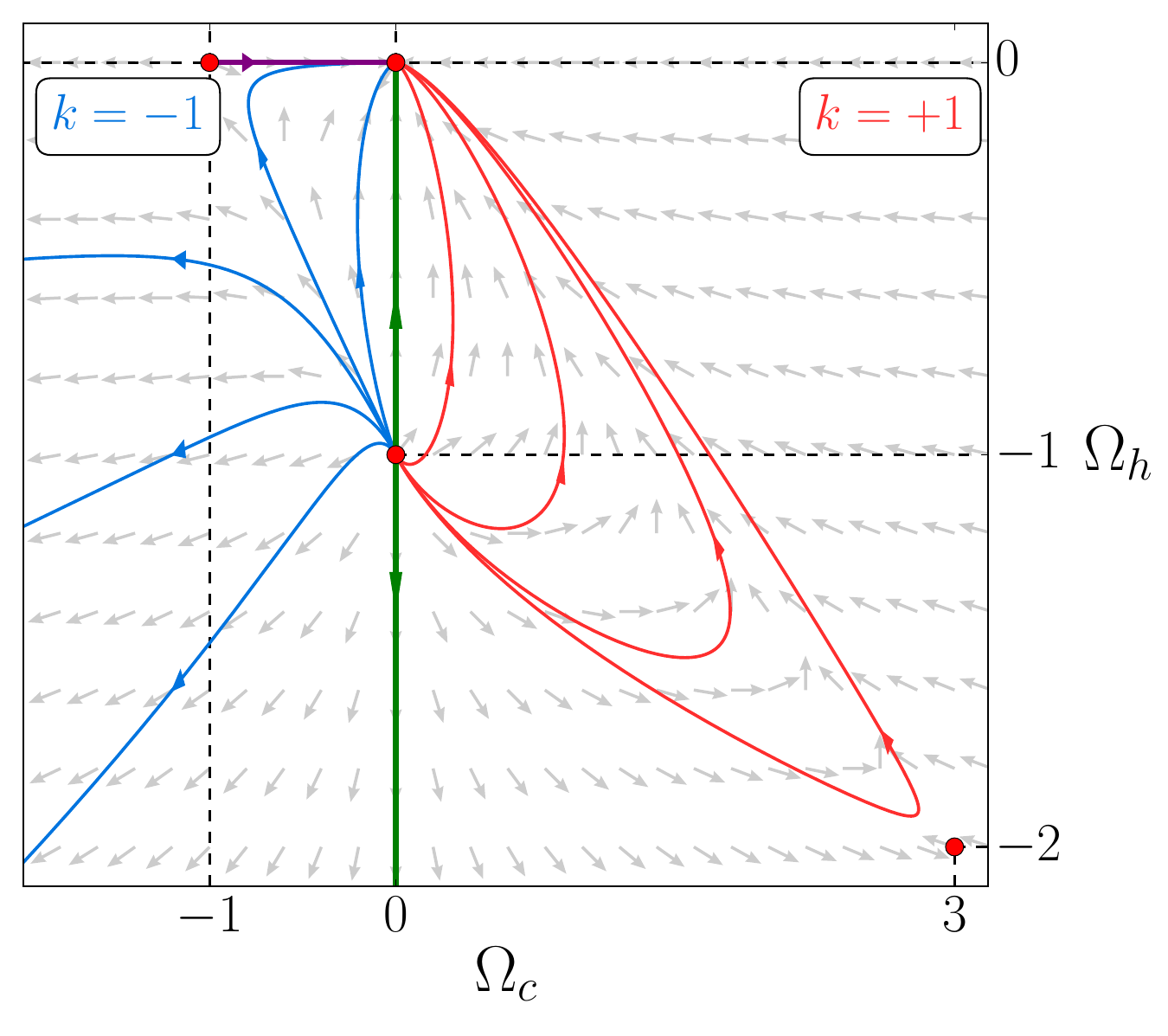} \\[-2mm]
		\small \textbf{(c)} $w=-1$, $w_5=-2$
		\label{fig:cur_ext_zerozero}
	\end{tabular}
	
	\vspace{1mm}
	
	\caption{Phase planes of the system in Eq.\autoref{eq:dynsys} for three sample sets of EoS parameters. Blue, green and red curves represent solutions in  open, flat and closed universes, respectively. Purple line in the third diagram is the inflationary solution in a four dimensional universe. In all diagrams solution curves below the line $\Omega_h + \Omega_c = -1$ are forbidden due to the fact that $\Omega_{\rho} < 0$ in those regions.}
	\label{fig:rad_lambda_phase}
\end{figure*}

While analyzing the fixed points, we come across the result that stabilization is possible at the expense of allowing a curvature dominated epoch along the time evolution of an open universe. However, the case does not apply equally to a closed universe due to the fact that all solutions diverge for $\Omega_c>0$ as seen in Fig.\autoref{fig:cur_ext_rad}.

To minimize the viable interval, the EoS parameters may be chosen such that they obey (\cite{bringmann1})
\begin{equation}
	3w+w_5=1
	\label{eq:line}
\end{equation}
on the $w-w_5$ plane, pointing at a highly relativistic fluid as one would expect to see, for instance, before the early inflation. Even more specifically, if we consider a completely isotropic universe in four spatial dimensions, we can choose the combination $w=w_5=1/4$ which is consistent with (\cite{bringmann1})
\begin{equation}
	p=\frac{\rho}{3+n}
\end{equation}
by noting $n=1$ for a single extra dimension. This choice leads to the evolutionary diagram in Fig.\autoref{fig:cur_ext_rad}. Unless an additional mechanism is introduced, radiation as the sole component here leads to a curvature dominated epoch for the case $\Omega_c < 0$. In other words, this phase ends with a negatively curved observed space and a stabilized extra dimension. 

Solution curves below the line $\Omega_h + \Omega_c = -1$ in Fig.\autoref{fig:cur_ext_rad} are forbidden due to the fact that $\Omega_{\rho} < 0$ in those regions violating the weak energy condition.

\vspace{2.5mm}

\subsection{The Range of EoS Parameters for Stabilization and Flatness}
Allowing $w_5<-1$ and $w<-1/3$, we succeed in obtaining a flat universe with a stabilized extra dimension as represented in Fig.\autoref{fig:cur_ext_zerozero} for a sample set of $w$,$w_5$. The curves emerge from $(\Omega_h,\Omega_c)=(-1,0)$, where both Hubble parameters are equal in magnitude, then move towards a negative-curvature dominated epoch while the fifth dimension shrinks in size.  Solution curves finally reach the point $(\Omega_h,\Omega_c)=(0,0)$ asymptotically in order to match the requirements of standard cosmology. Triangular region in Fig.\autoref{fig:cur_ext_zerozero} with convergent solutions for $k = -1$ exists invariably for the range of $w,w_5$ values in question. We are able to obtain this behavior through only a specific set of initial conditions obeying $\Omega_c + \Omega_h \geq -1 $ which also ensures that $\Omega_{\rho} \geq 0$.

As for the case of positively curved space, initial values that give convergent curves depend on EoS parameters since the flow alters due to the saddle point given as $(\Omega_c = 3, \Omega_h = -2)$ in Fig.\autoref{fig:cur_ext_zerozero}, that is, number of convergent solutions increase with decreasing $w,w_5$ values.

\subsection{An Invariant Manifold: Flat Universe}
Invariant manifolds are objects that divide phase spaces into distinct parts. They possess the feature of limiting the impact of an attractor to a region which is usually called the basin of attraction. In the presence of such manifold(s), a fixed point of the system cannot be a global attractor unless it appears on the (common intersection of) invariant manifold(s) in question.

In our system, the line $\Omega_c = 0$ which is illustrated with green solution curves in Fig.\autoref{fig:rad_lambda_phase} is a parameter-free invariant manifold representing a flat universe. One can predict this behavior even prior to the analysis of the dynamical system due to the fact that the curvature index $k$ in $\Omega_c$ term is a constant discrete parameter forbidding transitions between $\Omega_c < 0$ and $\Omega_c > 0$ cases. Namely, if the universe starts to evolve with a preferred curvature, it either stays exactly in the same state or becomes almost flat, i.e., no evolutionary path crosses the line with zero curvature.

On this invariant manifold Eq.\autoref{eq:dynsys} reduces to a one dimensional autonomous system described by the following differential equation
\begin{equation}
	\sfrac{d\Omega_h}{dN} = m(1+\Omega_h)(n+\Omega_h)
	\label{eq:flat_diff_eq}
\end{equation}
where $m=w_5-1$ and $n=(1+2w_5-3w)/(w_5-1)$. The fixed points of this one dimensional system are $-1$ and $-n$. However, this is an exactly solvable differential equation whose solution is
\begin{equation}
	\Omega_h = (n-1) \bigg[ 1 \mp \bigg(\! \sfrac{a}{a_o} \!\bigg)^{\!\!m(n-1)} \, \bigg]^{-1} - n 
	\label{eq:abs_val_eq}
\end{equation}
where $a_o$ is an integration constant. The plus sign applies only for the interval $\Omega_h \in (-1,1)$. Furthermore, writing $\Omega_h$ explicitly in terms of the scale factors and their derivatives, one obtains
\begin{equation}
	b(t) = \sfrac{b_o}{a_o^n} \bigg( \bigg[\! \sfrac{a(t)}{a_o} \!\bigg]^{-(1-w_5)} \mp \bigg[\! \sfrac{a(t)}{a_o} \!\bigg]^{1+2w_5-3w} \: \bigg)^{\!1/(1-w_5)}
	\label{eq:extdimscaleflat}
\end{equation}
where $b_o$ is another integration constant. Now assuming that $a(t)$ is an increasing function of time, let us examine the two special cases mentioned in the previous sections. For an isotropic fluid Eq.\autoref{eq:extdimscaleflat} becomes
\begin{equation}
	b(t) = b_o \, a_o \bigg( \bigg[\! \sfrac{a(t)}{a_o} \!\bigg]^{-(1-w_5)} \mp \bigg[\! \sfrac{a(t)}{a_o} \!\bigg]^{1-w_5} \: \bigg)^{\!1/(1-w_5)} \: .
	\label{eq:extdimscaleisotropicflat}
\end{equation}
For the range $\Omega_h \in (-1,1)$ and as $t \rightarrow \infty$, the scale factor $b(t)$ has the same rate of increment as $a(t)$, i.e. $b(t) \! \approx \! b_o \, a(t)$, provided that $w_5 < 1$. This occurs after an initial period of contraction. For $w_5 > 1$, $b(t)$ tends to zero independent of $w$. As for the range $\Omega_h \in (-\infty,-1) \cup (1,\infty)$ solutions of $b(t)$ alter severely with EoS parameters and due to the minus sign in the expression. There exist some unphysical solutions where $b(t)$ is either negative or imaginary. For instance, if we choose $w = w_5 = -1$, then $a(t)$ becomes limited. On the other hand, if we take $w = w_5 = 1/4$, then $b(t)$ increases linearly after a period of contraction similar to previous range of $\Omega_h$.

As for the case that satisfies both the compactification and the stabilization of the extra dimension with $1+2w_5-3w = 0$ the solution is
\begin{equation}
	b(t) = b_o\bigg( \bigg[\! \sfrac{a(t)}{a_o} \!\bigg]^{-(1-w_5)} \mp 1 \bigg)^{\!1/(1-w_5)}
	\label{eq:ext_scale_compact}
\end{equation}
which indicates that the final size of the fifth dimension turns out to be $b_o$ for $\Omega_h \in (-1,1)$ as $t \rightarrow \infty$. This expression gives a particular solution analyzed in (\cite{moham,bringmann1}) where the extra dimension is assumed to be stabilized. Another particular solution of the form $b=1/a^q$ was studied in (\cite{moham}) which is again the limiting case of Eq.\autoref{eq:ext_scale_compact} as $t \rightarrow 0$ and for $q=1$. For $\Omega_h \in (-\infty,-1) \cup (1,\infty)$ the situation is the same as the isotropic fluid case.

Now let us compare our results for a flat universe with the effect of the curvature by using phase planes given above. Looking at green curves in Fig.\autoref{fig:cur_ext_rad} and Fig.\autoref{fig:cur_ext_lambda}, we see identical flow directions both of which belong to an isotropic flat universe. On this invariant manifold, under the condition $1-w_5>0$, repeller and attractor points are seen to be $\Omega_h = -1$ (Point A) and $\Omega_h = 1$ (Point C), respectively. It is noteworthy to point out that the eigenvalues for these critical points in one-dimension are $\lambda = \pm 2(1-w_5)$ which also appear as the exponents in Eq.\autoref{eq:extdimscaleisotropicflat} to determine the time evolution of the scale factor $b(t)$. Depending on the characters of these fixed points, $b(t)$ will either tend to zero or become linearly proportional to $a(t)$ as mentioned before. On the other hand, if we change the curvature index from $k=0$ to $k=-1$, phase diagrams in figures become two dimensional and we obtain a different result for the parameters in Fig.\autoref{fig:cur_ext_rad} as the radiation component gives solutions in which the stabilization is achieved. In Fig.\autoref{fig:cur_ext_lambda}, the attractor remains unaltered and solution curves produce the same final picture as those of the invariant manifold for the case of a cosmological constant dominated universe. Thus, we see that while the stabilization of extra dimension cannot be achieved for an isotropic flat universe, it is possible to obtain such an outcome with the addition of the curvature term as long as $w > -1/3$.

As for the diagram in Fig.\autoref{fig:cur_ext_zerozero}, the attractor with $\Omega_h = 0$ implies that the solution (green curve) evolves to a state of stabilized extra dimension. If we use the condition $1 + 2w_5 - 3w =0$ together with $\lambda_2$ in Table \autoref{tab:1} for points A and C, we see that this result remains valid for $w,w_5 < 1$. For the cases where $w > 1$, $b(t)$ tends to zero as the flow direction alters due to the stability interchange of the critical points which can also be checked from Eq.\autoref{eq:ext_scale_compact}. Similar to the isotropic cases mentioned above, adding the curvature component to the system, we obtain solutions that realize both flatness and stabilization of extra dimension under the condition $w < -1/3$ and $w_5 < -1$. If the universe is already flat, then EoS parameters, for instance, can be $w_5=0$ and $w=1/3$ in order to obtain the desired solution. However, if the universe is not initially perfectly flat, then the condition on EoS parameters inevitably becomes $w < -1/3$ and $w_5 < -1$.

\section{CONCLUSION}
The idea that the (1+3)-dimensional universe we observe today could be part of a higher dimensional spacetime and could have emerged from it through a compactification mechanism seems to be quite an attractive approach. Indeed the cosmological evolution of extra dimensions can play a role in both the early and the late phases of the universe. For example in (\cite{Sahdev}) the idea of anisotropic extra dimensions is used to solve the horizon problem. Furthermore, the contribution of the cosmological evolution of extra dimensions to the dynamics of the universe is considered in literature to address also the current accelerated expansion of the universe (\cite{moham,Gu01,Gu03,pahwa11,Darabi,bringmann2}) and to account for the amount and nature of the dark matter by bringing the Kaluza-Klein tower of the particles (\cite{bringmann1,UED2}) to the stage. Mathematically all these attempts can work in some way or other because adding new degrees of freedom to the (1+3)-dimensional cosmological models by considering the evolution of extra dimensions allows one to obtain cosmologically viable scenarios. The fundamental coupling constants (for example $G$) may change with the size of the extra dimensional space and, thus, there are strong constraints coming from cosmological observations forcing that extra dimensions must be both compactified and stabilized before the nucleosynthesis (BBN).  In fact, in (\cite{Cline01}) some extra dimensional models are shown to be equaivalent to a class of Brans-Dicke theories which are observationally ruled out. There are still alternative ways of invoking extra dimensions to explain the accelerated expansion of the universe (\cite{2311,2311-2,2311-3,2320,2325,2325-2,2326}) by evading cosmological constraints with the use of additional ingredients. But the strategy that we follow in this paper is mostly different from all these approaches in that the main concern of our study is how to apply the idea of the presence and evolution of an extra dimensional model to the early universe.	

In this work we have considered the cosmological evolution of a five-dimensional universe as a simplified model of the early universe. The time evolution of the model is investigated in the framework of dynamical system analysis: all the critical points of the system are determined and their stability properties are examined according to the various ranges of EoS parameters. The main emphasis is on the classical stabilization of the extra dimension together with the time evolution of $\Omega_c$ of the three spatial dimensions, namely whether the flatness of the model be provided irrespective of its initial value. All these cosmologically relevant cases are summarized in Table \autoref{my-label}. An interesting observation is that although an isotropic perfect fluid in five-dimensional universe with equation of state parameters $w=w_5<-1/3$ can give rise to a flat universe, stabilization of the extra dimension is not possible in those cases. However, fluids with $w=w_5>-1/3$ (including the special case of radiation $w=w_5=1/4$) can lead to the stabilization of extra dimension and provide a negatively-curved three-dimensional space (Point B). The model, in this sense, can be issued for investigating the dynamics of the early universe. Furthermore, if the universe is assumed to be flat, then the only way to obtain stable solutions with a stabilized extra dimension is to impose the condition $1+2w_5-3w = 0$ which also works for the cases where $\Omega_c \neq 0$ as seen in Fig.\autoref{fig:cur_ext_zerozero}. Therefore, it appears that adding the curvature component to the model is necessary in order to achieve the stabilization of extra dimension with isotropic perfect fluids.

\section*{Acknowledgements}
This is a pre-print of an article published in \textit{Astrophysics and Space Science}. The final authenticated version is available online at: \url{https://doi.org/10.1007/s10509-018-3436-5}

\bibliographystyle{apsrev4-1}
\bibliography{references}

%merlin.mbs apsrev4-1.bst 2010-07-25 4.21a (PWD, AO, DPC) hacked
%Control: key (0)
%Control: author (72) initials jnrlst
%Control: editor formatted (1) identically to author
%Control: production of article title (-1) disabled
%Control: page (0) single
%Control: year (1) truncated
%Control: production of eprint (0) enabled
\providecommand{\noopsort}[1]{}\providecommand{\singleletter}[1]{#1}%
\begin{thebibliography}{31}%
\makeatletter
\providecommand \@ifxundefined [1]{%
 \@ifx{#1\undefined}
}%
\providecommand \@ifnum [1]{%
 \ifnum #1\expandafter \@firstoftwo
 \else \expandafter \@secondoftwo
 \fi
}%
\providecommand \@ifx [1]{%
 \ifx #1\expandafter \@firstoftwo
 \else \expandafter \@secondoftwo
 \fi
}%
\providecommand \natexlab [1]{#1}%
\providecommand \enquote  [1]{``#1''}%
\providecommand \bibnamefont  [1]{#1}%
\providecommand \bibfnamefont [1]{#1}%
\providecommand \citenamefont [1]{#1}%
\providecommand \href@noop [0]{\@secondoftwo}%
\providecommand \href [0]{\begingroup \@sanitize@url \@href}%
\providecommand \@href[1]{\@@startlink{#1}\@@href}%
\providecommand \@@href[1]{\endgroup#1\@@endlink}%
\providecommand \@sanitize@url [0]{\catcode `\\12\catcode `\$12\catcode
  `\&12\catcode `\#12\catcode `\^12\catcode `\_12\catcode `\%12\relax}%
\providecommand \@@startlink[1]{}%
\providecommand \@@endlink[0]{}%
\providecommand \url  [0]{\begingroup\@sanitize@url \@url }%
\providecommand \@url [1]{\endgroup\@href {#1}{\urlprefix }}%
\providecommand \urlprefix  [0]{URL }%
\providecommand \Eprint [0]{\href }%
\providecommand \doibase [0]{http://dx.doi.org/}%
\providecommand \selectlanguage [0]{\@gobble}%
\providecommand \bibinfo  [0]{\@secondoftwo}%
\providecommand \bibfield  [0]{\@secondoftwo}%
\providecommand \translation [1]{[#1]}%
\providecommand \BibitemOpen [0]{}%
\providecommand \bibitemStop [0]{}%
\providecommand \bibitemNoStop [0]{.\EOS\space}%
\providecommand \EOS [0]{\spacefactor3000\relax}%
\providecommand \BibitemShut  [1]{\csname bibitem#1\endcsname}%
\let\auto@bib@innerbib\@empty
%</preamble>
\bibitem [{\citenamefont {Kaluza}(1921)}]{Kaluza}%
  \BibitemOpen
  \bibfield  {author} {\bibinfo {author} {\bibfnamefont {T.}~\bibnamefont
  {Kaluza}},\ }\href@noop {} {\bibfield  {journal} {\bibinfo  {journal}
  {Sitzungsber. Preuss. Akad. Wiss. Berlin (Math. Phys.)}\ }\textbf {\bibinfo
  {volume} {1921}},\ \bibinfo {pages} {966} (\bibinfo {year}
  {1921})}\BibitemShut {NoStop}%
\bibitem [{\citenamefont {Klein}(1926)}]{Klein}%
  \BibitemOpen
  \bibfield  {author} {\bibinfo {author} {\bibfnamefont {O.}~\bibnamefont
  {Klein}},\ }\href@noop {} {\bibfield  {journal} {\bibinfo  {journal} {Z.
  Physik}\ }\textbf {\bibinfo {volume} {37}},\ \bibinfo {pages} {69} (\bibinfo
  {year} {1926})}\BibitemShut {NoStop}%
\bibitem [{\citenamefont {{Goenner}}(2004)}]{goenner04}%
  \BibitemOpen
  \bibfield  {author} {\bibinfo {author} {\bibfnamefont {H.~F.~M.}\
  \bibnamefont {{Goenner}}},\ }\href@noop {} {\bibfield  {journal} {\bibinfo
  {journal} {Living Reviews in Relativity}\ }\textbf {\bibinfo {volume} {7}}
  (\bibinfo {year} {2004})}\BibitemShut {NoStop}%
\bibitem [{\citenamefont {{Randall}}\ and\ \citenamefont
  {{Sundrum}}(1999{\natexlab{a}})}]{rs1}%
  \BibitemOpen
  \bibfield  {author} {\bibinfo {author} {\bibfnamefont {L.}~\bibnamefont
  {{Randall}}}\ and\ \bibinfo {author} {\bibfnamefont {R.}~\bibnamefont
  {{Sundrum}}},\ }\href@noop {} {\bibfield  {journal} {\bibinfo  {journal}
  {Physical Review Letters}\ }\textbf {\bibinfo {volume} {83}},\ \bibinfo
  {pages} {4690} (\bibinfo {year} {1999}{\natexlab{a}})},\ \Eprint
  {http://arxiv.org/abs/hep-th/9906064} {arXiv:hep-th/9906064} \BibitemShut
  {NoStop}%
\bibitem [{\citenamefont {{Randall}}\ and\ \citenamefont
  {{Sundrum}}(1999{\natexlab{b}})}]{rs2}%
  \BibitemOpen
  \bibfield  {author} {\bibinfo {author} {\bibfnamefont {L.}~\bibnamefont
  {{Randall}}}\ and\ \bibinfo {author} {\bibfnamefont {R.}~\bibnamefont
  {{Sundrum}}},\ }\href@noop {} {\bibfield  {journal} {\bibinfo  {journal}
  {Physical Review Letters}\ }\textbf {\bibinfo {volume} {83}},\ \bibinfo
  {pages} {3370} (\bibinfo {year} {1999}{\natexlab{b}})},\ \Eprint
  {http://arxiv.org/abs/hep-ph/9905221} {arXiv:hep-ph/9905221} \BibitemShut
  {NoStop}%
\bibitem [{\citenamefont {{Arkani-Hamed}}\ \emph {et~al.}(1998)\citenamefont
  {{Arkani-Hamed}}, \citenamefont {{Dimopoulos}},\ and\ \citenamefont
  {{Dvali}}}]{add}%
  \BibitemOpen
  \bibfield  {author} {\bibinfo {author} {\bibfnamefont {N.}~\bibnamefont
  {{Arkani-Hamed}}}, \bibinfo {author} {\bibfnamefont {S.}~\bibnamefont
  {{Dimopoulos}}}, \ and\ \bibinfo {author} {\bibfnamefont {G.}~\bibnamefont
  {{Dvali}}},\ }\href@noop {} {\bibfield  {journal} {\bibinfo  {journal}
  {Physics Letters B}\ }\textbf {\bibinfo {volume} {429}},\ \bibinfo {pages}
  {263} (\bibinfo {year} {1998})},\ \Eprint
  {http://arxiv.org/abs/hep-ph/9803315} {hep-ph/9803315} \BibitemShut {NoStop}%
\bibitem [{\citenamefont {{Dvali}}\ \emph {et~al.}(2000)\citenamefont
  {{Dvali}}, \citenamefont {{Gabadadze}},\ and\ \citenamefont
  {{Porrati}}}]{dgp}%
  \BibitemOpen
  \bibfield  {author} {\bibinfo {author} {\bibfnamefont {G.}~\bibnamefont
  {{Dvali}}}, \bibinfo {author} {\bibfnamefont {G.}~\bibnamefont
  {{Gabadadze}}}, \ and\ \bibinfo {author} {\bibfnamefont {M.}~\bibnamefont
  {{Porrati}}},\ }\href@noop {} {\bibfield  {journal} {\bibinfo  {journal}
  {Physics Letters B}\ }\textbf {\bibinfo {volume} {485}},\ \bibinfo {pages}
  {208} (\bibinfo {year} {2000})},\ \Eprint
  {http://arxiv.org/abs/hep-th/0005016} {arXiv:hep-th/0005016} \BibitemShut
  {NoStop}%
\bibitem [{\citenamefont {{Maartens}}(2004)}]{maartens}%
  \BibitemOpen
  \bibfield  {author} {\bibinfo {author} {\bibfnamefont {R.}~\bibnamefont
  {{Maartens}}},\ }\href@noop {} {\bibfield  {journal} {\bibinfo  {journal}
  {Living Reviews in Relativity}\ }\textbf {\bibinfo {volume} {7}},\ \bibinfo
  {eid} {7} (\bibinfo {year} {2004})},\ \Eprint
  {http://arxiv.org/abs/gr-qc/0312059} {arXiv:gr-qc/0312059} \BibitemShut
  {NoStop}%
\bibitem [{\citenamefont {Perlmutter}\ \emph {et~al.}(1998)\citenamefont
  {Perlmutter} \emph {et~al.}}]{perlmutter}%
  \BibitemOpen
  \bibfield  {author} {\bibinfo {author} {\bibfnamefont {S.}~\bibnamefont
  {Perlmutter}} \emph {et~al.},\ }\href@noop {} {\bibfield  {journal} {\bibinfo
   {journal} {Nature}\ }\textbf {\bibinfo {volume} {391}},\ \bibinfo {pages}
  {51} (\bibinfo {year} {1998})},\ \Eprint
  {http://arxiv.org/abs/astro-ph/9712212} {arXiv:astro-ph/9712212} \BibitemShut
  {NoStop}%
\bibitem [{\citenamefont {Riess}\ \emph {et~al.}(1998)\citenamefont {Riess}
  \emph {et~al.}}]{riess}%
  \BibitemOpen
  \bibfield  {author} {\bibinfo {author} {\bibfnamefont {A.~G.}\ \bibnamefont
  {Riess}} \emph {et~al.},\ }\href@noop {} {\bibfield  {journal} {\bibinfo
  {journal} {The Astronomical Journal}\ }\textbf {\bibinfo {volume} {116}},\
  \bibinfo {pages} {1009} (\bibinfo {year} {1998})},\ \Eprint
  {http://arxiv.org/abs/astro-ph/9805201} {arXiv:astro-ph/9805201} \BibitemShut
  {NoStop}%
\bibitem [{\citenamefont {{Townsend}}\ and\ \citenamefont
  {{Wohlfarth}}(2003)}]{town03}%
  \BibitemOpen
  \bibfield  {author} {\bibinfo {author} {\bibfnamefont {P.~K.}\ \bibnamefont
  {{Townsend}}}\ and\ \bibinfo {author} {\bibfnamefont {M.~N.}\ \bibnamefont
  {{Wohlfarth}}},\ }\href@noop {} {\bibfield  {journal} {\bibinfo  {journal}
  {Physical Review Letters}\ }\textbf {\bibinfo {volume} {91}},\ \bibinfo {eid}
  {061302} (\bibinfo {year} {2003})},\ \Eprint
  {http://arxiv.org/abs/hep-th/0303097} {arXiv:hep-th/0303097} \BibitemShut
  {NoStop}%
\bibitem [{\citenamefont {Mohammedi}(2002)}]{moham}%
  \BibitemOpen
  \bibfield  {author} {\bibinfo {author} {\bibfnamefont {N.}~\bibnamefont
  {Mohammedi}},\ }\href@noop {} {\bibfield  {journal} {\bibinfo  {journal}
  {Phys. Rev. D}\ }\textbf {\bibinfo {volume} {65}},\ \bibinfo {pages} {104018}
  (\bibinfo {year} {2002})},\ \Eprint {http://arxiv.org/abs/hep-th/0202119}
  {arXiv:hep-th/0202119} \BibitemShut {NoStop}%
\bibitem [{\citenamefont {{Pahwa}}\ \emph {et~al.}(2011)\citenamefont
  {{Pahwa}}, \citenamefont {{Choudhury}},\ and\ \citenamefont
  {{Seshadri}}}]{pahwa11}%
  \BibitemOpen
  \bibfield  {author} {\bibinfo {author} {\bibfnamefont {I.}~\bibnamefont
  {{Pahwa}}}, \bibinfo {author} {\bibfnamefont {D.}~\bibnamefont
  {{Choudhury}}}, \ and\ \bibinfo {author} {\bibfnamefont {T.~R.}\ \bibnamefont
  {{Seshadri}}},\ }\href@noop {} {\bibfield  {journal} {\bibinfo  {journal}
  {\jcap}\ }\textbf {\bibinfo {volume} {9}},\ \bibinfo {eid} {015} (\bibinfo
  {year} {2011})},\ \Eprint {http://arxiv.org/abs/1104.1925} {arXiv:1104.1925
  [gr-qc]} \BibitemShut {NoStop}%
\bibitem [{\citenamefont {{Appelquist}}\ and\ \citenamefont
  {{Dobrescu}}(2001)}]{appel01}%
  \BibitemOpen
  \bibfield  {author} {\bibinfo {author} {\bibfnamefont {T.}~\bibnamefont
  {{Appelquist}}}\ and\ \bibinfo {author} {\bibfnamefont {B.~A.}\ \bibnamefont
  {{Dobrescu}}},\ }\href@noop {} {\bibfield  {journal} {\bibinfo  {journal}
  {Physics Letters B}\ }\textbf {\bibinfo {volume} {516}},\ \bibinfo {pages}
  {85} (\bibinfo {year} {2001})},\ \Eprint
  {http://arxiv.org/abs/hep-ph/0106140} {arXiv:hep-ph/0106140} \BibitemShut
  {NoStop}%
\bibitem [{\citenamefont {{Bringmann}}\ \emph {et~al.}(2003)\citenamefont
  {{Bringmann}}, \citenamefont {{Eriksson}},\ and\ \citenamefont
  {{Gustafsson}}}]{bringmann1}%
  \BibitemOpen
  \bibfield  {author} {\bibinfo {author} {\bibfnamefont {T.}~\bibnamefont
  {{Bringmann}}}, \bibinfo {author} {\bibfnamefont {M.}~\bibnamefont
  {{Eriksson}}}, \ and\ \bibinfo {author} {\bibfnamefont {M.}~\bibnamefont
  {{Gustafsson}}},\ }\href@noop {} {\bibfield  {journal} {\bibinfo  {journal}
  {\prd}\ }\textbf {\bibinfo {volume} {68}},\ \bibinfo {eid} {063516} (\bibinfo
  {year} {2003})},\ \Eprint {http://arxiv.org/abs/astro-ph/0303497}
  {arXiv:astro-ph/0303497} \BibitemShut {NoStop}%
\bibitem [{\citenamefont {Bahamonde}\ \emph {et~al.}(2017)\citenamefont
  {Bahamonde}, \citenamefont {Boehmer}, \citenamefont {Carloni}, \citenamefont
  {Copeland}, \citenamefont {Fang},\ and\ \citenamefont
  {Tamanini}}]{dynreview}%
  \BibitemOpen
  \bibfield  {author} {\bibinfo {author} {\bibfnamefont {S.}~\bibnamefont
  {Bahamonde}}, \bibinfo {author} {\bibfnamefont {C.~G.}\ \bibnamefont
  {Boehmer}}, \bibinfo {author} {\bibfnamefont {S.}~\bibnamefont {Carloni}},
  \bibinfo {author} {\bibfnamefont {E.~J.}\ \bibnamefont {Copeland}}, \bibinfo
  {author} {\bibfnamefont {W.}~\bibnamefont {Fang}}, \ and\ \bibinfo {author}
  {\bibfnamefont {N.}~\bibnamefont {Tamanini}},\ }\href@noop {} {\  (\bibinfo
  {year} {2017})},\ \Eprint {http://arxiv.org/abs/1712.03107} {arXiv:1712.03107
  [gr-qc]} \BibitemShut {NoStop}%
%%CITATION = ARXIV:1712.03107;%%
\bibitem [{\citenamefont {{Carroll}}\ \emph {et~al.}(2002)\citenamefont
  {{Carroll}}, \citenamefont {{Geddes}}, \citenamefont {{Hoffman}},\ and\
  \citenamefont {{Wald}}}]{wald-carroll02}%
  \BibitemOpen
  \bibfield  {author} {\bibinfo {author} {\bibfnamefont {S.~M.}\ \bibnamefont
  {{Carroll}}}, \bibinfo {author} {\bibfnamefont {J.}~\bibnamefont {{Geddes}}},
  \bibinfo {author} {\bibfnamefont {M.~B.}\ \bibnamefont {{Hoffman}}}, \ and\
  \bibinfo {author} {\bibfnamefont {R.~M.}\ \bibnamefont {{Wald}}},\
  }\href@noop {} {\bibfield  {journal} {\bibinfo  {journal} {\prd}\ }\textbf
  {\bibinfo {volume} {66}},\ \bibinfo {eid} {024036} (\bibinfo {year}
  {2002})},\ \Eprint {http://arxiv.org/abs/hep-th/0110149} {hep-th/0110149}
  \BibitemShut {NoStop}%
\bibitem [{\citenamefont {{Bringmann}}\ and\ \citenamefont
  {{Eriksson}}(2003)}]{bringmann2}%
  \BibitemOpen
  \bibfield  {author} {\bibinfo {author} {\bibfnamefont {T.}~\bibnamefont
  {{Bringmann}}}\ and\ \bibinfo {author} {\bibfnamefont {M.}~\bibnamefont
  {{Eriksson}}},\ }\href@noop {} {\bibfield  {journal} {\bibinfo  {journal}
  {\jcap}\ }\textbf {\bibinfo {volume} {10}},\ \bibinfo {eid} {006} (\bibinfo
  {year} {2003})},\ \Eprint {http://arxiv.org/abs/astro-ph/0308498}
  {arXiv:astro-ph/0308498} \BibitemShut {NoStop}%
\bibitem [{\citenamefont {Georgalas}\ \emph {et~al.}(2017)\citenamefont
  {Georgalas}, \citenamefont {Karydas},\ and\ \citenamefont
  {Papantonopoulos}}]{UED2}%
  \BibitemOpen
  \bibfield  {author} {\bibinfo {author} {\bibfnamefont {B.~C.}\ \bibnamefont
  {Georgalas}}, \bibinfo {author} {\bibfnamefont {S.}~\bibnamefont {Karydas}},
  \ and\ \bibinfo {author} {\bibfnamefont {E.}~\bibnamefont
  {Papantonopoulos}},\ }\href@noop {} {\  (\bibinfo {year} {2017})},\ \Eprint
  {http://arxiv.org/abs/1711.02723} {arXiv:1711.02723 [gr-qc]} \BibitemShut
  {NoStop}%
%%CITATION = ARXIV:1711.02723;%%
\bibitem [{\citenamefont {Sahdev}(1984)}]{Sahdev}%
  \BibitemOpen
  \bibfield  {author} {\bibinfo {author} {\bibfnamefont {D.}~\bibnamefont
  {Sahdev}},\ }\href@noop {} {\bibfield  {journal} {\bibinfo  {journal} {Phys.
  Rev. D}\ }\textbf {\bibinfo {volume} {30}},\ \bibinfo {pages} {2495}
  (\bibinfo {year} {1984})}\BibitemShut {NoStop}%
\bibitem [{\citenamefont {Gu}\ and\ \citenamefont {Hwang}(2002)}]{Gu01}%
  \BibitemOpen
  \bibfield  {author} {\bibinfo {author} {\bibfnamefont {J.-A.}\ \bibnamefont
  {Gu}}\ and\ \bibinfo {author} {\bibfnamefont {W.-Y.~P.}\ \bibnamefont
  {Hwang}},\ }\href@noop {} {\bibfield  {journal} {\bibinfo  {journal} {Phys.
  Rev. D}\ }\textbf {\bibinfo {volume} {66}},\ \bibinfo {pages} {024003}
  (\bibinfo {year} {2002})},\ \Eprint {http://arxiv.org/abs/astro-ph/0112565}
  {arXiv:astro-ph/0112565} \BibitemShut {NoStop}%
\bibitem [{\citenamefont {Gu}\ \emph {et~al.}(2004)\citenamefont {Gu},
  \citenamefont {Hwang},\ and\ \citenamefont {Tsai}}]{Gu03}%
  \BibitemOpen
  \bibfield  {author} {\bibinfo {author} {\bibfnamefont {J.-A.}\ \bibnamefont
  {Gu}}, \bibinfo {author} {\bibfnamefont {W.-Y.}\ \bibnamefont {Hwang}}, \
  and\ \bibinfo {author} {\bibfnamefont {J.-W.}\ \bibnamefont {Tsai}},\
  }\href@noop {} {\bibfield  {journal} {\bibinfo  {journal} {Nuclear Physics
  B}\ }\textbf {\bibinfo {volume} {700}},\ \bibinfo {pages} {313 } (\bibinfo
  {year} {2004})},\ \Eprint {http://arxiv.org/abs/astro-ph/0403641}
  {arXiv:astro-ph/0403641} \BibitemShut {NoStop}%
\bibitem [{\citenamefont {Darabi}(2003)}]{Darabi}%
  \BibitemOpen
  \bibfield  {author} {\bibinfo {author} {\bibfnamefont {F.}~\bibnamefont
  {Darabi}},\ }\href@noop {} {\bibfield  {journal} {\bibinfo  {journal}
  {Classical and Quantum Gravity}\ }\textbf {\bibinfo {volume} {20}},\ \bibinfo
  {pages} {3385} (\bibinfo {year} {2003})},\ \Eprint
  {http://arxiv.org/abs/gr-qc/0301075} {arXiv:gr-qc/0301075} \BibitemShut
  {NoStop}%
\bibitem [{\citenamefont {Cline}\ and\ \citenamefont {Vinet}(2003)}]{Cline01}%
  \BibitemOpen
  \bibfield  {author} {\bibinfo {author} {\bibfnamefont {J.~M.}\ \bibnamefont
  {Cline}}\ and\ \bibinfo {author} {\bibfnamefont {J.}~\bibnamefont {Vinet}},\
  }\href@noop {} {\bibfield  {journal} {\bibinfo  {journal} {Phys. Rev. D}\
  }\textbf {\bibinfo {volume} {68}},\ \bibinfo {pages} {025015} (\bibinfo
  {year} {2003})},\ \Eprint {http://arxiv.org/abs/hep-ph/0211284}
  {arXiv:hep-ph/0211284} \BibitemShut {NoStop}%
\bibitem [{\citenamefont {Deffayet}\ \emph {et~al.}(2002)\citenamefont
  {Deffayet}, \citenamefont {Dvali},\ and\ \citenamefont {Gabadadze}}]{2311}%
  \BibitemOpen
  \bibfield  {author} {\bibinfo {author} {\bibfnamefont {C.}~\bibnamefont
  {Deffayet}}, \bibinfo {author} {\bibfnamefont {G.}~\bibnamefont {Dvali}}, \
  and\ \bibinfo {author} {\bibfnamefont {G.}~\bibnamefont {Gabadadze}},\
  }\href@noop {} {\bibfield  {journal} {\bibinfo  {journal} {Phys. Rev. D}\
  }\textbf {\bibinfo {volume} {65}},\ \bibinfo {pages} {044023} (\bibinfo
  {year} {2002})},\ \Eprint {http://arxiv.org/abs/astro-ph/0105068}
  {arXiv:astro-ph/0105068} \BibitemShut {NoStop}%
\bibitem [{\citenamefont {Sahni}\ and\ \citenamefont {Shtanov}(2003)}]{2311-2}%
  \BibitemOpen
  \bibfield  {author} {\bibinfo {author} {\bibfnamefont {V.}~\bibnamefont
  {Sahni}}\ and\ \bibinfo {author} {\bibfnamefont {Y.}~\bibnamefont
  {Shtanov}},\ }\href@noop {} {\bibfield  {journal} {\bibinfo  {journal}
  {Journal of Cosmology and Astroparticle Physics}\ }\textbf {\bibinfo {volume}
  {2003}},\ \bibinfo {pages} {014} (\bibinfo {year} {2003})},\ \Eprint
  {http://arxiv.org/abs/astro-ph/0202346} {arXiv:astro-ph/0202346} \BibitemShut
  {NoStop}%
\bibitem [{\citenamefont {Maia}\ \emph {et~al.}(2004)\citenamefont {Maia},
  \citenamefont {Monte},\ and\ \citenamefont {Maia}}]{2311-3}%
  \BibitemOpen
  \bibfield  {author} {\bibinfo {author} {\bibfnamefont {M.}~\bibnamefont
  {Maia}}, \bibinfo {author} {\bibfnamefont {E.~M.}\ \bibnamefont {Monte}}, \
  and\ \bibinfo {author} {\bibfnamefont {J.}~\bibnamefont {Maia}},\ }\href@noop
  {} {\bibfield  {journal} {\bibinfo  {journal} {Physics Letters B}\ }\textbf
  {\bibinfo {volume} {585}},\ \bibinfo {pages} {11 } (\bibinfo {year}
  {2004})},\ \Eprint {http://arxiv.org/abs/astro-ph/0208223}
  {arXiv:astro-ph/0208223} \BibitemShut {NoStop}%
\bibitem [{\citenamefont {Perivolaropoulos}\ and\ \citenamefont
  {Sourdis}(2002)}]{2320}%
  \BibitemOpen
  \bibfield  {author} {\bibinfo {author} {\bibfnamefont {L.}~\bibnamefont
  {Perivolaropoulos}}\ and\ \bibinfo {author} {\bibfnamefont {C.}~\bibnamefont
  {Sourdis}},\ }\href@noop {} {\bibfield  {journal} {\bibinfo  {journal} {Phys.
  Rev. D}\ }\textbf {\bibinfo {volume} {66}},\ \bibinfo {pages} {084018}
  (\bibinfo {year} {2002})},\ \Eprint {http://arxiv.org/abs/hep-ph/0204155}
  {arXiv:hep-ph/0204155} \BibitemShut {NoStop}%
\bibitem [{\citenamefont {Brax}\ \emph
  {et~al.}(2003{\natexlab{a}})\citenamefont {Brax}, \citenamefont {van~de
  Bruck}, \citenamefont {Davis},\ and\ \citenamefont {Rhodes}}]{2325}%
  \BibitemOpen
  \bibfield  {author} {\bibinfo {author} {\bibfnamefont {P.}~\bibnamefont
  {Brax}}, \bibinfo {author} {\bibfnamefont {C.}~\bibnamefont {van~de Bruck}},
  \bibinfo {author} {\bibfnamefont {A.-C.}\ \bibnamefont {Davis}}, \ and\
  \bibinfo {author} {\bibfnamefont {C.~S.}\ \bibnamefont {Rhodes}},\
  }\href@noop {} {\bibfield  {journal} {\bibinfo  {journal} {Phys. Rev. D}\
  }\textbf {\bibinfo {volume} {67}},\ \bibinfo {pages} {023512} (\bibinfo
  {year} {2003}{\natexlab{a}})},\ \Eprint {http://arxiv.org/abs/hep-th/0209158}
  {arXiv:hep-th/0209158} \BibitemShut {NoStop}%
\bibitem [{\citenamefont {Brax}\ \emph
  {et~al.}(2003{\natexlab{b}})\citenamefont {Brax}, \citenamefont {van~de
  Bruck}, \citenamefont {Davis},\ and\ \citenamefont {Rhodes}}]{2325-2}%
  \BibitemOpen
  \bibfield  {author} {\bibinfo {author} {\bibfnamefont {P.}~\bibnamefont
  {Brax}}, \bibinfo {author} {\bibfnamefont {C.}~\bibnamefont {van~de Bruck}},
  \bibinfo {author} {\bibfnamefont {A.-C.}\ \bibnamefont {Davis}}, \ and\
  \bibinfo {author} {\bibfnamefont {C.}~\bibnamefont {Rhodes}},\ }\href@noop {}
  {\bibfield  {journal} {\bibinfo  {journal} {Astrophysics and Space Science}\
  }\textbf {\bibinfo {volume} {283}},\ \bibinfo {pages} {627} (\bibinfo {year}
  {2003}{\natexlab{b}})},\ \Eprint {http://arxiv.org/abs/hep-ph/0210057}
  {arXiv:hep-ph/0210057} \BibitemShut {NoStop}%
\bibitem [{\citenamefont {Binétruy}\ \emph {et~al.}(2001)\citenamefont
  {Binétruy}, \citenamefont {Deffayet},\ and\ \citenamefont
  {Langlois}}]{2326}%
  \BibitemOpen
  \bibfield  {author} {\bibinfo {author} {\bibfnamefont {P.}~\bibnamefont
  {Binétruy}}, \bibinfo {author} {\bibfnamefont {C.}~\bibnamefont {Deffayet}},
  \ and\ \bibinfo {author} {\bibfnamefont {D.}~\bibnamefont {Langlois}},\
  }\href@noop {} {\bibfield  {journal} {\bibinfo  {journal} {Nuclear Physics
  B}\ }\textbf {\bibinfo {volume} {615}},\ \bibinfo {pages} {219 } (\bibinfo
  {year} {2001})},\ \Eprint {http://arxiv.org/abs/hep-th/0101234}
  {arXiv:hep-th/0101234} \BibitemShut {NoStop}%
\end{thebibliography}%

\end{document}